\documentclass[runningheads]{llncs}
\usepackage[T1]{fontenc}
\usepackage{graphicx}
\usepackage[misc]{ifsym}
\begin{document}
\title{Hidden or Formal Architects: Understanding Who Makes Architectural Decisions in Practice}
\titlerunning{Who Makes Architectural Decisions in Practice}
%
\author{Klara Borowa\inst{1}\textsuperscript{\Letter}\orcidID{0000-0002-7160-5950} \and
Dominik Nowak\inst{1}\orcidID{0009-0008-5453-0150} \and
Mikołaj Nowak\inst{1}\orcidID{0009-0000-3781-3329}}
\authorrunning{Borowa et al.}

\institute{Warsaw University of Technology, Institute of Control and Computation Engineering, 
Warsaw, Poland \\
\email{klara.borowa@pw.edu.pl}}

\maketitle              
\begin{abstract}
 
Empirical research on software architecture sometimes focuses on individuals holding the formal title of software architect. 
However, not all companies have individuals hired in such roles.
Despite no formal architects, all systems have architectures, and there must be practitioners making the architectural decisions.
This study aims to identify who, in practice, makes architectural decisions and in what environments the existence of a formal architect is perceived as necessary.
This research employed a method consisting of a questionnaire with 54 participants from different companies and seven follow-up interviews. 
The findings indicate that architectural decisions are often made by individuals without a formal architect title. 
A formal architect is perceived as essential mainly in large companies and large teams. 
The results of this study show that many practitioners taking part in ADM may have been omitted by previous research, particularly if it focused on formal architects.

\keywords{Software Architecture \and Architectural decision-making \and Agile \and Informal Architect}
\end{abstract}

\section{Introduction}
The concept of software architecture as a set of architectural decisions (AD) was introduced by Jansen and Bosch~\cite{jansen2005software}, and has shaped much research since.
Farenhorst et al.~\cite{farenhorst2009lonesome} focused on ADs in their large-scale survey from 2006, and found that software architects were lonesome decision-makers who neglected sharing architectural knowledge.
However, by 2015, based on an interview study with participants from 22 companies, Groher and Weinreich~\cite{groher2015study} revealed that architectural decision-making (ADM) was typically a group effort with one dedicated team member having the final say. 
The shift towards group ADM coincides with the rise of agile software development.  
In 2023, around 71\% of practitioners reported working in an agile or partially agile environment~\cite{stateofagile18}. 
Yet, in agile, a software architect's role is considered ambiguous~\cite{yang2016systematic}, e.g., Scrum assumes that no architect role exists, and that the team collectively makes decisions~\cite{Schwaber2020}.

Software architecture researchers often gather data from practitioners~\cite{razavian2019empirical}. Often, this research only focuses on the perspectives of architects~\cite{de2023let}, or more experienced developers~\cite{paris2023impact}. 
However, due to the dominance of agile methods, the people who perform ADM are likely not to have the architect title. 
As such, when searching for ``architects'', researchers may risk obtaining data that is biased towards practitioners and teams that do have a formal architect.
For instance, Demir et al.~\cite{demir2024factors} surveyed 101 practitioners recruited from software-architecture communities and found that ADM is typically performed by teams of around 5 people, comprising architects and software developers. Yet, due to their sampling strategy, practitioners who were not formal architects or were not actively interested in software architecture were naturally excluded from the study.

Still, every software system has an architecture, whether its creators explicitly address ADM or let it ``emerge'' from everyday development. 
This makes it unclear which practitioners we should sample to obtain a complete picture of ADM phenomena. Additionally, sampling only formal architects, i.e., practitioners explicitly assigned to that role, may lead to overlooking certain kinds of companies and teams that view such a dedicated role as unnecessary.
    \begin{enumerate}
        \item \textbf{RQ1:} Who makes architectural decisions in practice?
        \item \textbf{RQ2:} In what environments is a formal software architect necessary?

    \end{enumerate}

To achieve this goal, we performed a two-step mixed-methods empirical inquiry: through a questionnaire with 54 participants from a variety of companies, and 7 follow-up interviews to deeper explore the results of the questionnaire.

Overall, we found a great variety in companies' approaches to architectural decision-making depending on many factors. These include the findings that: (1) Most teams have no  formal architect,
    (2) Practitioners of all experience levels usually take part in ADM, (3) Agile teams can be very different regarding having a formal architect, (4) A formal architect is perceived as more necessary in large teams, larger companies, fully on-site teams, and teams that already have one.

\section{Related Work} \label{sec:2-rel-work}
In this section, we describe how the software architect's role has been viewed and researched over the years, existing research on who makes architectural decisions in various environments, and the current practices related to sampling in software architecture research.

\textbf{Software architect's role.}
In 1999, Kruchten~\cite{kruchten1999software} described the role and skill set of a software architect and a software architecture team. He later expands this~\cite{kruchten2008software} with an exploration of typical tasks and responsibilities associated with the software architect's role.
Later, through a large survey of 279 practicing architects, Farenhorst et al.~\cite{farenhorst2009lonesome}, assessed the typical architectural activities and methods for sharing architectural knowledge. They found that while the main activity of an architect was architectural decision-making, the least was documenting their decisions. Overall, Farenhorst et al. described the role of an architect as a ``lonesome'' with less emphasis on communication with other practitioners.
This was followed by Sherman and Unkelos-Shpigel~\cite{sherman2014software} who noticed architects did not perceive their role in accordance with existing software architecture literature -- particularly, architects did not believe that they are responsible for communication between the development team and business stakeholders. Sherman and Hadar~\cite{sherman2015toward} expanded this research by defining the soft aspects of an architect's role.



\textbf{Software architect's role in agile software development.}
As of 2003, around 71\% of software practitioners use agile methodologies to some extent~\cite{stateofagile18}. Some of these, like Scrum~\cite{Schwaber2020}, define a set of roles for the software development team that does not include an architect. Some, like Kanban~\cite{Meyer2014}, do not strictly define any roles. Others, like SAFe~\cite{safe2025}, explicitly state the importance and define the roles of system and enterprise architects.
Abrahamsson et al.~\cite{abrahamsson2010agility} noted that agile practices and the traditional approach to software architecture appear to be contradictory, yet agile developers perceive software architecture as important for their work.
In the same year, Breivold et al.~\cite{breivold2010does} performed a systematic review and showed a lack of research on the connection between architecture and agile.
 Later, efforts to make architecture co-exist with agile were made~\cite{babar2014making}, including researching the role of architects in agile teams~\cite{maric2016role} -- where architects made the last decision based on group discussion. Researchers also focused on challenges faced by architects in Scrum teams~\cite{angelov2016architects}, and proposed group decision-making approaches~\cite{lopes2017architectural}.


Yet, as shown in the systematic mapping by Yang et al. performed in 2016, ``The role of architects in agile development is not clear, and the agile architects have far fewer standard practices to follow.''~\cite{yang2016systematic}. 
No systematic mapping nor review has been published on this topic since then, to the best of the authors' knowledge.


\textbf{Architectural decision-making in different environments.}
There seems to have been a change since the ``lonely architect'' concept was found by Farenhorst et al.~\cite{farenhorst2009lonesome}.
LaToza et al.~\cite{latoza2013study} performed an interview study and found that architectural decision-making was a social process. 
Griher and Weinereich~\cite{groher2015study} later performed an interview study with 25 software architects, team leads, and senior developers from 22 different companies. They found that architectural alternatives are largely discussed by whole teams -- yet usually one dedicated person had the final say. They also found that low-level decisions can be taken by developers.
Britto et al.~\cite{britto2016software} reported a case study from Ericsson, showcasing a large and complex distributed project that had to include a team of architects. 
Rekha and Muccini~\cite{v_group_2018} researched group decision-making practices based on a questionnaire with 35 participants -- it showed that most practitioners made architectural decisions in groups composed of 3-5 people.
Paris and Guerra~\cite{paris2023impact} presented a case study regarding the impact of work in a start-up company on architectural decisions. In that context, they found that some decisions are made by smaller teams (if they are viewed as impacting only that team), while decisions important in the greater context were made by Teach Leaders.
De Almeida and van der Hoek~\cite{de2023let} performed a mixed-methods study about whiteboard architecture meetings. They found that practitioners of various levels of experience were viewed as important for these meetings. Particularly, they found that novices are viewed as necessary since they were not biased by previous experiences.

All of these studies showcase that the landscape of architectural decision-making is extremely rich, and depending on specific situations, different people with different levels of knowledge and experience can take part and provide valuable input in the process.
However, no recent study (since Farenhorst et al.~\cite{farenhorst2009lonesome} that discovered on ``lonesome architects'') so far has attempted to showcase who makes architectural decisions in a larger sample than 26 companies~\cite{v_group_2018} -- and none focused on exploring if this is dependent on such factors as software development methodology, experience, team size, company size, and the location of team members. The closest study is the one performed by Demir et al.~\cite{demir2024factors} surveyed 101 practitioners about factors affecting architectural decision‐making process. However, the focus on that study was not on who makes decisions and the author purposefully sought practitioners explicitly taking part in software architecture online communities - which may have biased their sample towards companies that employ formal architects. 

\textbf{Sampling and software architecture research.}
Empirical software engineering draws inspiration from medical research~\cite{kitchenham2004evidence}, where sample sizes are supposed to represent target populations. 
However, software engineering differs from medicine in that there is a clear lack of information on target populations~\cite{fernandez2019empirical}.
Razavian et al.~\cite{razavian2019empirical} showcase typical participants of software architecture studies. These are,  in 69\% of studies, practitioners and in 5\%, students and practitioners. Razavian et al. also note that in most cases, the sampling method is not specified.
As such, the sampled populations in software architecture studies can vary significantly. Some focus only on senior practitioners~\cite{weinreich2015expert}, while others focus only on architects~\cite{farenhorst2009lonesome}. Rarely do researchers attempt to include practitioners with various experience levels~\cite {borowa2023rationales}.

In this study, we explore who makes architectural decisions based on experience, role, company size, team size, and software development methodology. While our sample is not enough to define a target population, our study is a pointer towards what groups of architectural decision-makers may have been underrepresented in previous research, and who researchers should include in future studies. For example, since we found that the role of a formal architect is mainly considered necessary in larger companies, by researching only formal architects, small companies may be omitted.

\section{Method} \label{sec:3-method}

We applied a mixed‑methods approach with an online questionnaire with follow‑up interviews (see \textbf{replication package}~\cite{additional_material}).
As such, we involved methodological triangulation,
unlike work that relied solely on interviews \cite{groher2015study} or questionnaires~\cite{demir2024factors}. 

\subsection{Questionnaire}
\noindent\textbf{Design: }
Our questionnaire consisted mainly of closed-ended questions arranged thematically with the following: demographics, software development methodology, ADM, and architect's role.

We asked participants to leave an email address if they agreed to a follow‑up interview. 

\noindent\textbf{Data gathering: }
We distributed the questionnaire at two IT career fairs at Warsaw Univerity of Technology and \textit{via} social media (LinkedIn, Facebook). We avoided channels dominated by one company to ensure organizational diversity.

\noindent\textbf{Data analysis: }
We analyzed the ordinal-scale responses with descriptive statistics, including mean, median, and interquartile range.
Then, we compared groups regarding five factors: formal architect presence, software development methodology, company size, team size, and place of work.

\subsection{Interview}
\noindent\textbf{Design: }
The interview questions were based on patterns and anomalies identified in the questionnaire data that related to our research questions. 

Additionally, we added a few participant-specific questions for noteworthy cases, e.g., for a respondent who self-identified as an architect but rarely made ADs.

\noindent\textbf{Data gathering: }
Twelve questionnaire respondents provided consent for a follow-up interview. We invited all of them, but only seven responded. Each interview was held entirely online, recorded, and transcribed by the interviewer. 
Each interview lasted about an hour. At the start, we explained the study, planned data use, and obtained explicit informed consent.

\noindent\textbf{Data analysis: }
We analyzed interview transcripts using structural and open coding~\cite{saldana2021coding}.
The structural codes linked RQs with particular transcript fragments. Open codes captured named concepts that addressed the RQs.

Two authors developed and iteratively refined the coding scheme. Iteratively, one coded the transcript, and the other independently reviewed it. When new codes emerged, one coder checked the old transcripts' coding to establish if these concepts were missed or mislabeled in a previous iteration. The authors held two coding refinement meetings to improve the organization of the coding scheme.

\subsection{Sample}
\begin{figure}
    \centering

    \includegraphics[width=0.6\linewidth]{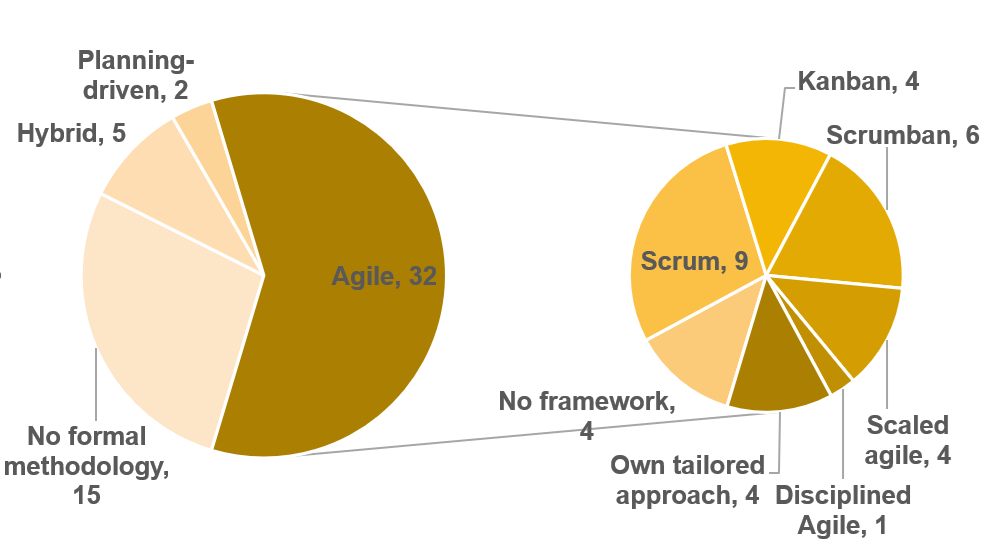}
    \caption{Participants' software development methodologies}
    \label{fig:methodologies}
    \vspace{-15pt}    
\end{figure}

Overall, 54 participants from different companies filled out the questionnaire. Most self-identified as developers (39) or tech leads (5), with only 2 having the architect title. They had diverse levels of experience and worked in companies of various sizes in 20 different domains. \textbf{In 38 of 54 teams, no formal architect role existed.} Most respondents reported using agile or hybrid software development methods (Figure~\ref{fig:methodologies}). Seven of the questionnaire participants, all from different domains and company sizes, took part in the follow‑up interviews. Full participant demographics are available in our replication package~\cite{additional_material}.

\section{Results} \label{sec:4-results}

In this section, we present the results obtained by analyzing the questionnaires and interviews. 

\subsection{Questionnaire} \label{res-questionary}

\begin{figure}
    \centering
    \includegraphics[width=0.8\linewidth]{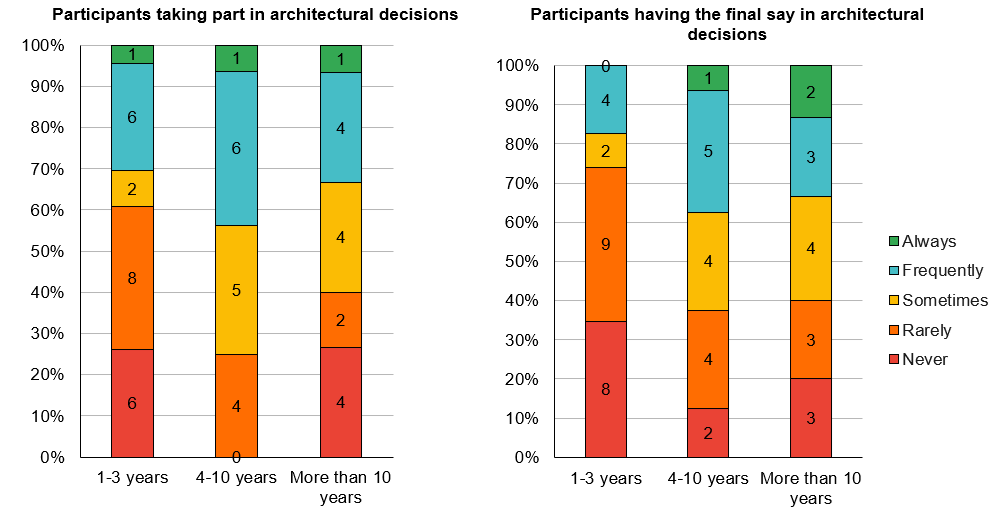}
    \caption{Architectural decision-making by experience}
    \label{fig:who_exp}
\end{figure}

\noindent\textbf{Experience.}
Figure~\ref{fig:who_exp} showcases who makes architectural decisions depending on experience level. This if based on our participants' demographics and whether they choose the ``Me'' answers to two specific questions: (1) ``Who typically makes those architectural decisions in your team?'' and (2) ``Who has the final say in those decisions?''.

We divided the results by experience level to explore whether individuals with a specific experience level make more architectural decisions: juniors (1-3 years), middles (4-10 years), and seniors (over 10 years). Since we expected that group decision-making may take place, participants reported both participation and having the authority of making the final architectural decision.

\begin{figure*}
    \centering
    \includegraphics[width=\linewidth]{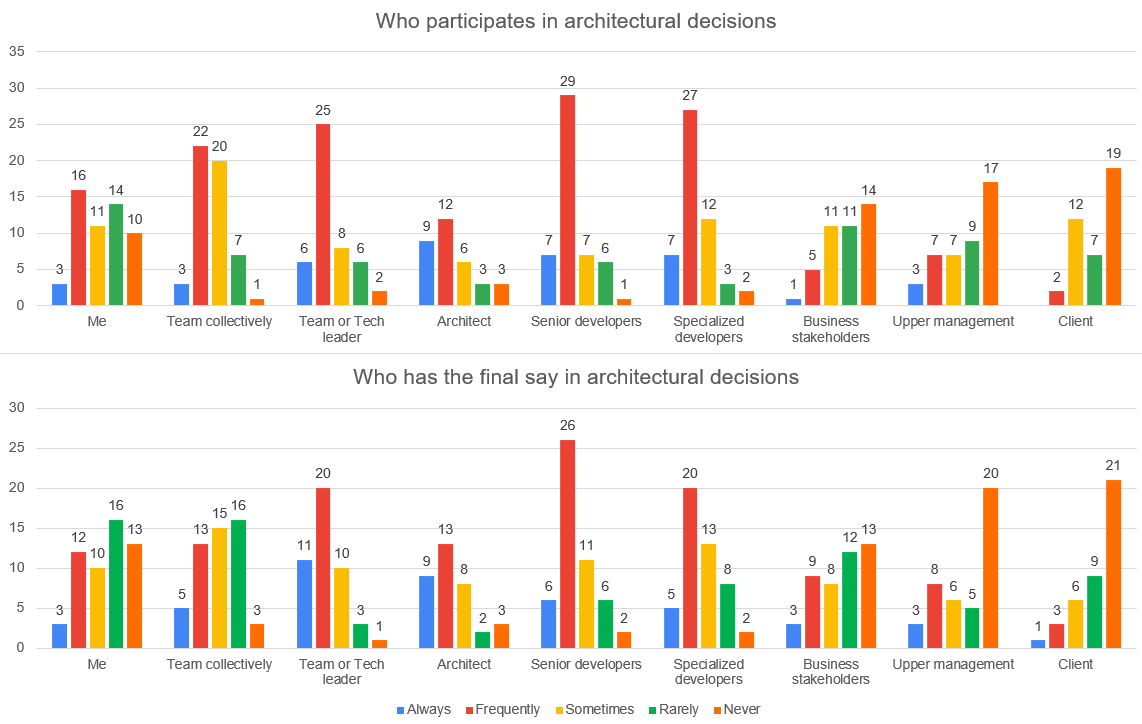}
    \caption{Architectural decision-making by role}
    \label{fig:who_role}
\end{figure*} 

Among juniors, 26\% reported never taking part in architectural decisions. However, only 30\% do so frequently or always -- which means that juniors overall take part in decisions, but less often. In contrast, none of the middles indicated that they never participated in such decisions -- and nearly 40\% reported making architectural decisions frequently or always, while only 25\% did so rarely. 
Notably, seniors exhibited a pattern similar to middles but with a slight shift toward less frequent decision-making. 
This trend prompted further investigation, and the findings from related interview findings are presented in Section \ref{res-interview}. 

In the case of having the final say in architectural decisions, the pattern seems similar to participation. However, the proportion of respondents reporting having the authority for final decisions was notably smaller -- particularly in the case of juniors and middles. Seniors, however, reported having the final decision power more often than they actually participated in decision-making, which means that they have more power over decisions than they usually make use of.

\noindent\textbf{Role.}
Figure~\ref{fig:who_role} also showcases who, in the participants' companies, usually takes part in architectural decisions, and who has the authority for final decisions. It is based on how our participants responded the questions, for many roles: (1) ``Who typically makes those architectural decisions in your team?'' and (2) ``Who has the final say in those decisions?''.

Overall, the dominating influence in both of these categories seems to belong to senior developers (i.e., the ones with the most experience),  specialized developers (i.e., the ones with the most domain knowledge on the particular topic), and team/technical leaders (i.e., formal leaders of the team with technical backgrounds). 

However, it should be noted that quite frequently, the teams discuss the decision collectively, although the final decision is less likely to depend on the whole team's agreement. As such, it seems that decision-making is usually a group activity with some figure of authority having the final say. 

Architects, when they are present, participate and have a deciding role. 
However, another notable finding emerges from these results. More participants marked that an architect frequently or always takes part in decision-making (21) than reported having a formal architect in their team (16). This leads to the conclusion that there most likely exists such a role as an \textbf{informal architect} - i.e., an architect that is not formally employed as an architect, yet their coworkers perceive them as the architect. We explored this issue further in the interviews.

Finally, it seems that business stakeholders, upper management, and clients usually do not take part in architectural decisions. However, it is clearly not always the case since in the case of each of these roles, at least one participant noted that they always participate in architectural decisions. As such, their influence cannot be completely overlooked.

\noindent\textbf{Software Development Methodology.}
In this section, we explore whether the presence of a formal architect is related to the software development methodology used by a team. 

\begin{table}
\caption{Architects depending on software development methodology.}
\scalebox{0.9}{\begin{tabular}{|p{2.2cm}|p{2cm}|p{3cm}|p{3cm}|p{3cm}|} 
\hline
\textbf{Methodology} & \textbf{Answer} & \textbf{Formal} & \textbf{Enterprise } & \textbf{Considers } \\
\textbf{} & \textbf{count} & \textbf{ architect} & \textbf{architect}& \textbf{architect }  \\
\textbf{} & \textbf{} & \textbf{present} & \textbf{present}& \textbf{necessary }  \\
\textbf{} & \textbf{} & \textbf{} & \textbf{}& \textbf{(median) }  \\
\hline
No formal 	&	&	&	&\\
methodology	&	15	&	13.33\%&	26.68\%	&2\\
\hline
Agile: 	&	&	&	&\\
Scrum	&	9	&	11.11\% &	11.12\%	&2\\
\hline
Agile: 	&	&	&	&\\
Scrumban	&	6	&	33.33\%	&	\textbf{66.68\%}	&\textbf{3.5}\\
\hline
Hybrid	&	5	&	20\%	&	\textbf{60\%}	&3\\
\hline
Agile: 	&	&	&	&\\
Own tailored	&	4	&	\textbf{75\%	}&	\textbf{50\%}	&\textbf{4}\\
\hline
Agile: 	&	&	&	&\\
Scaled agile	&	4	&	\textbf{75\%}	&\textbf{75\%}	&\textbf{3.5}\\
\hline
Agile: 	&	&	&	&\\
No framework	&	4&	\textbf{50\%}	&	\textbf{100\%}	&3\\
\hline
Agile: 	&	&	&	&\\
Kanban	&	4	&	25\%	&	25\%	&3\\
\hline
Planning-driven	&	2	&	\textbf{50\%}	&	0\%	&2\\
\hline
Agile: 	&	&	&	&\\
Disciplined Agile	&	1	&	0\%	&	\textbf{100\%}	&\textbf{5}\\

\hline
\end{tabular}}

\label{tab:method_vs_architect}
\end{table}

Table \ref{tab:method_vs_architect} showcases: (1) The amount of participants using a particular methodology, (2) The percentage of them that have a formal architect in their team, (3) The percentage of them that have an enterprise architect in their company and, (4) Their median rating of whether they consider the presence of an architect necessary (on a 1-5 scale from ``unnecessary'' to ``very necessary'').

Overall, it is worth noting that enterprise architects were present more often than team-level architects. As such, it seems that it is more typical to have a formal architect working on the enterprise level.

Teams working with no agile practices and no formal framework at all seem to rarely have any kind of formal architect present, and do not consider them to be necessary.
This contrasts with the approach of teams that use agile practices but do not employ a specific agile framework: all of them had an enterprise-level architect, and half of them had a team-level architect.

Teams using planning-driven approaches, although less represented in our study (only 2 participants), sometimes had a formal team-level architect but no enterprise architect. They also did not consider an architect's presence necessary.

Teams using hybrid approaches usually had an enterprise architect (60\%) but were less likely to have a team-level one (20\%). They also had a neutral outlook on an architect's necessity.

Overall, the presence or absence of an architect in agile methodologies is very different depending on the methodology. It seems that both team-level and enterprise architects are employed less frequently in Scrum and Kanban, but are more commonly present in other agile approaches. Scrum practitioners also usually consider an architect to be unnecessary, but other agile practitioners are either neutral or positive about an architect's necessity.

\begin{figure*}
    \centering
    \includegraphics[width=1\linewidth]{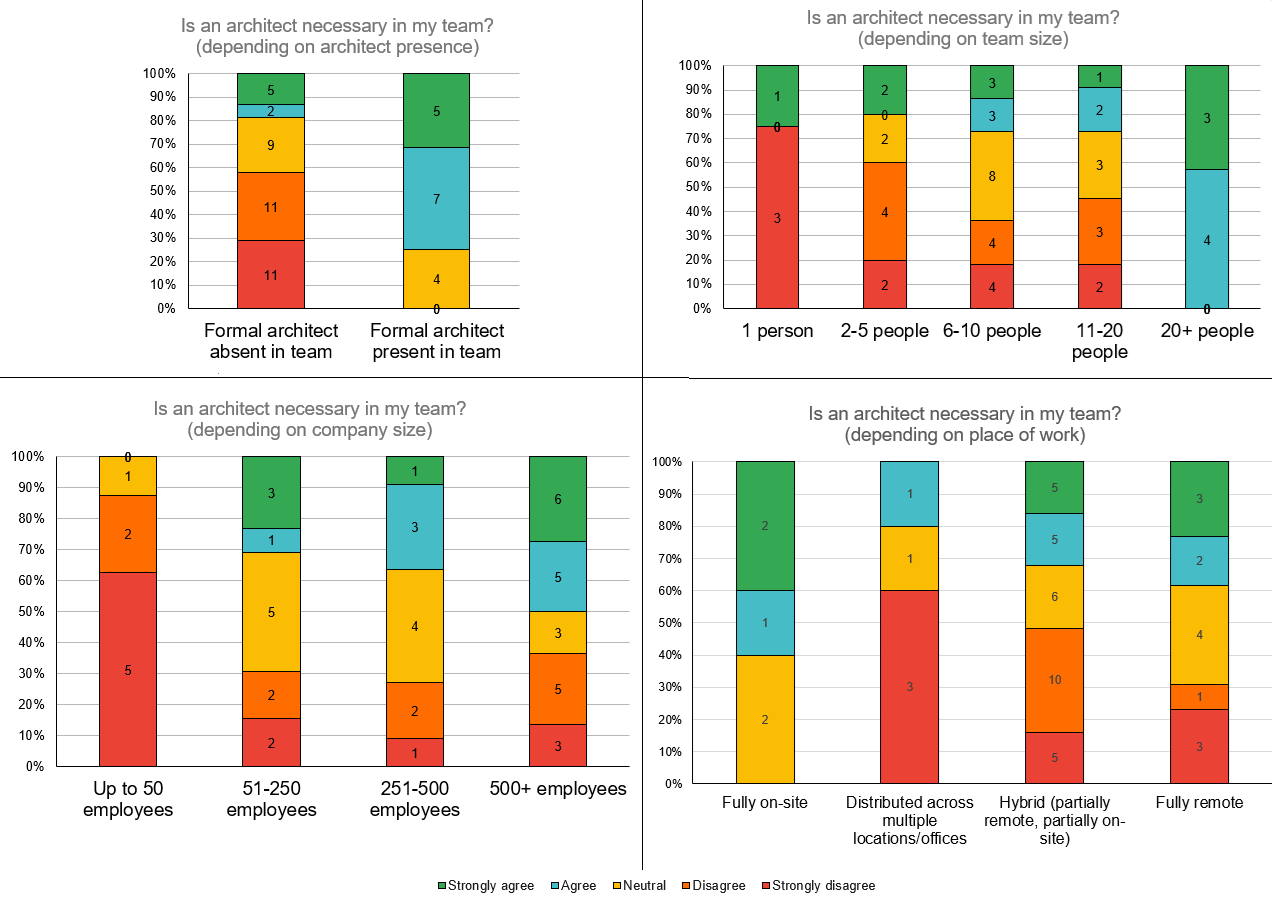}
    \caption{Evaluation of formal architect necessity}
    \label{fig:necessity}
\end{figure*}

\noindent\textbf{Perception of formal architect's necessity.}
Figure~\ref{fig:necessity} showcases how practitioners view the necessity of an architect in their team depending on various factors. We based this evaluation on the participants' answer to the question ``I think a formal architect is essential in projects I take part in'' which could be answered on a scale from 1 (Strongly Disagree) to 5(Strongly Agree).

Firstly, when a formal architect is actually present in the team, in no case were they considered unnecessary, and most participants considered the architect to be important. However, when no architect was present, only a small part of the participants considered that they lacked an architect in their team.

Secondly, team size seems to be possibly the most important factor connected to the necessity of an architect. For big teams (over 20 people), all participants agreed that an architect was necessary. The responses were very different for mid-sized teams (6-20 people), but most participants in small teams (1-5 people) considered an architect to be unnecessary.

Thirdly, company size also seems to be connected to architect necessity perceptions. The difference is not as extreme as in cases of team size, but it is clear that while in small companies (up to 50 employees), architects are usually considered unnecessary, half of the teams in big companies (over 500 employees) consider an architect to be necessary. This trend seems to match with mid-sized companies being in the middle of this trend. 

Finally, the connection between the place of work and the perception of architectural necessity turned out to be very divisive. Only in the case of teams working fully on-site did all participants perceive an architect to be necessary or were neutral. In all other cases (various offices, hybrid, and full remote), at least 3 participants strongly disagreed, and the answers varied. The least necessity for a formal architect was recorded from teams distributed across different offices.

\subsection{Interview results} \label{res-interview}

\subsubsection{Who makes architectural decisions?}
The interviews confirm that architectural decisions are \textbf{most often made by experienced individuals}, regardless of their official role or title. 
Interviewee 7 explained this, ``there is a kind of consensus that the most experienced people have the mandate to make these kinds of decisions, and the greatest chance of making the right ones.'' 
In general, experience was consistently stated as the primary factor in determining who should make architectural decisions. 
%
Additionally, \textbf{collaborative decision-making} was highlighted, particularly in agile settings, where interviewee 5 observed, ``[In Agile] this is distributed more across the team (...)''.

Another recurring answer in the interviews was the emergence of \textbf{informal architects} in teams without a formally designated role. Participants described how the most experienced or domain-knowledgeable team members often naturally assumed architectural responsibilities. As Interviewee 6 put it, ``In most cases, the person who becomes the [informal] architect is the one with the most experience in the given project''. 
These individuals are typically led by initiative and peer recognition, rather than a formal title, highlighting a practice-driven, trust-based approach to architectural leadership. This informal role is often temporary and context-dependent, with team members stepping into architectural responsibilities during the planning or design phases and returning to development work thereafter. 
As Interviewee 6 described, ''In a self-organizing team, you take on the architectural role during system design, but once [...] are done, you go back to being a regular developer''. 

Additionally, in teams without a formally designated architect, \textbf{a Tech Leader often fills the gap}. As Interviewee 7 pointed out, “There is often someone called a Tech Lead, and you could say that this person effectively acts as the architect, even though they are not formally named as one.”

In larger organizations, enterprise architects were frequently identified as influential figures in high-level decision structures. Our participants described \textbf{a hierarchical model,} where routine decisions are delegated to development teams, while strategic or critical decisions remained centralized. As Interviewee 6 explained, ``strategic, top-level architectural decisions are made by [enterprise] architects [...] and come to the team as settled decisions [...], while lower-level decisions concerning smaller components are made by the team''. 

The hierarchical model is also related to \textbf{architectural decision delegation} that we noticed during the interviews, i.e., that middle-level practitioners handled more architectural decisions than seniors, because seniors were busier with other tasks. Interviewee 3 describes this: ``Rather, such things [decisions] are delegated, perhaps to other people who are also very experienced, but do not have as much seniority, and I have the impression that those who have been there for more than 10 years mainly have, let's say, a representative function.''

Furthermore, \textbf{clients} occasionally seem to take part in architectural decisions, which is usually not perceived favorably by practitioners. Interviewee 7 expressed this clearly, stating, ``Sometimes the client doesn't want the best solution,'' while also acknowledging that, ''Technically or architecturally, the decision may not be optimal, but from the client's perspective, it is''.

Finally, it is worth noting that \textbf{junior developers} often avoid taking on architectural responsibilities, mostly out of fear or uncertainty. As Interviewee 1 noted, ``A young developer in the company is probably a bit too nervous to write any kind of documentation or propose architectural designs''.

\subsubsection{In what environment is a formal architect necessary?}
The presence of a formal architect was seen as \textbf{essential in large-scale or complex projects}, where integration, coordination, and planning must be closely managed. As stated by Interviewee 1: ``If you have a small company, then you probably also have a smaller project, so the architecture may have less of an impact. If you have a huge project with hundreds of people working on it, then an architect would be useful.'' 

Participants also stated that \textbf{a remote or distributed work model does not} necessarily require formal architectural oversight to maintain alignment across locations, as Interviewee 6 said: ``Work that simply needs to be done will get done regardless of whether it’s in-person or remote. A good architect doesn’t need to see you face-to-face to present the architecture''.

Similarly, in \textbf{projects focusing on maintenance,} a formal architect was often considered unnecessary, as existing team members could effectively manage decisions, as Interviewee 2 stated: ``Right now, we're mostly maintaining and updating what already exists. And when we do develop something, it's usually a minor system—so a formal architect would be unnecessary, as there wouldn't be much for them to do''. 

Additionally, the perceived need for an architect was often connected with \textbf{respondents' past experiences}, reflecting a form of bias in how participants evaluated architectural roles. As Interviewee 2 explained, ``It often stems from habit—people working in teams without an architect get used to making decisions on their own. [...] But once such a role is introduced, they probably wouldn't want to work without one anymore''.

\subsubsection{The impact of a formal architect's presence/absence.}
Interviewees linked the presence of a formal architect with \textbf{improved risk management, and reduced technical debt}, as stated by Interviewee 2: ``[The architect] based on his experience, he figured out that combining this or that isn't a good idea, or, for example, it's cool that you want to use some cool new framework, but from experience, it can be risky because it's new. (...) And then you had to maintain a dead program from scratch, or write it from scratch in a different framework. Which at this point is simply technical debt.''

Architects were also seen as \textbf{enablers of cross-team communication}, i.e. Interviewee 3 stated: ``(...) the architect is responsible, among other things, for establishing communication between design teams and between their projects.''. 

Additionally, an architect's presence was also viewed as \textbf{beneficial for strategic and purchasing} decisions within the organization, as Interviewee 4 said: ``in the long run, they will save about 10 million ''.

Furthermore, a formal architect was believed to speed up the team's overall work. For example, Interviewee 3 stated: ``The architect has a more intuitive, perhaps broader perspective on the entire system, so in all these sprints, when opinions need to be given or something needs to be changed, they are simply a source of knowledge, and that is why they can speed up decision-making at these meetings, which tend to drag on.''

One participant noted that enterprise and team-level architects may have conflicting roles with each other. However, more usually, participants reported \textbf{complementary roles between enterprise and team-level architects}, especially when responsibilities were clearly defined. As Interviewee 6 noted, ``They actually have different tasks—different scopes and levels of abstraction they work on,'' highlighting how a clear separation of concerns can support effective collaboration between architectural roles.

\section{Discussion} \label{sec:5-dis}

\noindent\textbf{RQ1: Who makes architectural decisions in practice?}
Our study reveals that architectural decisions are not made exclusively by formally designated architects -- formal architects are actually part of only around a third of software development teams. This does not align with the previous findings of Demir et al.~\cite{demir2024factors}, who found that only 5 of 101 participants no architect on their projects -- suggesting that our sampling, by approaching practitioners at random at events and in non-architecture online spaces, yielded extremely different results. 

Regarding experience, practitioners of all levels usually have some experience in ADM. Only 26\% of junior practitioners reported never taking part in architectural decisions. 
However, the final say in decisions often was in the hands of senior developers, specialized developers, and tech leaders. As such, it seems that usually, ADM is a group process with various team members (which does fit the findings of Demir et al.~\cite{demir2024factors}), with the most experienced developers or Tech Leads assuming the majority of decision-making responsibilities. However, our key finding is that this ``final say'' decision-makers can be considered as informal architects, who assume this role through social legitimacy and peer recognition.

These findings introduce important implications for researchers regarding sampling in software architecture -- or rather, the areas that may have been overlooked.
By gathering only data from formal architects, researchers are very likely to omit data from even two-thirds of possible companies. As such, tech leaders and senior developers should also be included in samples more prominently than formal architects.

Additionally, since practitioners with 3-10 years of experience take part in most decisions, more mid-career practitioners should be represented in samples. 
Junior practitioners also seem to participate in many cases, so they should not be omitted either.

Regarding software development methodology, the study shows a quite diverse picture for each agile method. 
This means that generalizing ADM practices for all agile methods cannot be made based on a small subset of them.

\noindent\textbf{RQ2: In what environments is the presence of a formal software architect necessary?}
Firstly, the results suggest that formal architects are seen as essential primarily in large teams and large companies.
 
As such, it seems that the necessity for a formal architect scales with project complexity and organization size.
Secondly, architects are more valued in fully on-site teams, whereas interview insights suggested their role may be even more critical in remote settings. This disparity most likely results from architects' perceived visibility in co-located settings, where casual conversations and immediate assistance are more noticeable. In contrast, remote teams may underappreciate the architect's work. 

This holds crucial implications for researchers. When sampling for ADM research, focusing on formal architects may omit small companies and small teams.

\section{Threats to validity} \label{sec:6-thr-val}

\textbf{Construct Validity:}
A key threat in our study comes from the subjective interpretation of core concepts by participants, e.g. on what an architectural decision is. 
To mitigate this threat, before asking who was involved in making architectural decisions, we presented a predefined list of possible decisions. 
Additionally, participants may have responded based on their distorted memory or beliefs.
To address these ambiguities and distortions, we included the follow-up interviews.  
This triangulation allowed us to cross-validate the questionnaire responses.

\textbf{Internal Validity:}
One threat is the researchers' bias, since during qualitative interview analysis, there was the threat that a single researcher may unconsciously search for data confirming their beliefs. To avoid this, we applied separate coding by two authors, followed by a discussion to reach an agreement.

\textbf{External Validity:}
A primary threat is the relatively small sample size and its limited geographical diversity. Most questionnaire responses came from participants based in Poland and other European countries. This limits our ability to generalize the findings.
However, we cross-checked our sample with existing literature stating that 71\% of practitioners worked in an agile or partially-agile companies~\cite{stateofagile18}, and in the case of our sample, it was 69\%, which we believe may be an indicator that our sample has similar proportions to the real-world state.
Additionally, regarding the interviewees, since these were conducted among respondents who provided consent, there is a risk that this leads to a selection bias where individuals with outspoken feelings mainly provided the data.

\section{Conclusion} \label{sec:7-conc}

This study focuses on who makes architectural decisions in practice in software development teams and in what environment a formal architect is considered necessary. To explore this, we gathered data through a questionnaire and interviews.

Our key findings are that architectural decisions are frequently made by individuals who do not hold the formal architect title. 
Decisions are made either by whole teams or Tech Leads and senior developers who commonly step into informal architect roles. However, teams generally recognize the value of having a dedicated architect in large-scale or complex projects, where strategic oversight, responsibility, and coordination are crucial.

Implications \textbf{for researchers} are related to sampling strategies for software architecture research. Since it seems that often all team members take part in decisions their perspectives should not be excluded from architectural decision-making studies. 
Additionally, researchers should ensure that they avoid excluding samples from small companies, which may be less visible.  
\textbf{For practitioners},  our results highlight that many empirical findings on software architecture may be biased towards larger organizations and teams that employ formal architects. When using such results, practitioners should critically assess whether their own context matches the contexts represented in the specific empirical study.

\textbf{Future Work} could focus on: (1) Replicating this study in the context, particularly in non-European settings, to provide more generalizable results, (2) Creating sampling guidelines specific to software architecture research, (3) Exploring the role of previously hidden architectural decision-makers - particularly the junior practitioners that sometimes make architectural decisions.

\textbf{Data availability: } We provide the participant demographic data, the questionnaire, the interview plan, and a list of qualitative codes)~\cite{additional_material}. We do not provide raw questionnaire answers or transcripts since we did not obtain explicit consent to do so.


%
%
%
\bibliographystyle{splncs04}
\bibliography{references}
%




\end{document}